%% file: main.tex
\documentclass{article}
\usepackage[T1]{fontenc}
\usepackage{spconf,amsmath,graphicx, caption}
\usepackage{amssymb, gensymb}
\usepackage{tikz, pgfplots}
\usetikzlibrary{positioning, arrows.meta, calc, positioning, quotes}
\usepackage{fancyhdr, ragged2e}

\fancypagestyle{firstpage}{
  \fancyhf{} %
  \fancyfoot[C]{\fontsize{9pt}{9pt}\selectfont \justify{© 2023 IEEE. Personal use of this material is permitted. Permission from IEEE must be obtained for all other uses, in any current or future media, including reprinting/republishing this material for advertising or promotional purposes, creating new collective works, for resale or redistribution to servers or lists, or reuse of any copyrighted component of this work in other works.}}
}

\usepackage[font=small,skip=2pt]{caption}
\newcommand{\mysection}[1]{
  \vspace{-5pt}
  \section{#1}
  \vspace{-5pt}
}

\newcommand{\mysubsection}[1]{
  \vspace{-7pt}
  \subsection{#1}
  \vspace{-5pt}
}

\title{A Real-Time Active Speaker Detection System Integrating an Audio-Visual Signal with a Spatial Querying Mechanism}

\address{Author Affiliation(s)}
\name{\emph{Ilya Gurvich, Ido Leichter, Dharmendar Reddy Palle, Yossi Asher, Alon Vinnikov,} \\ 
      \emph{Igor Abramovski,  Vishak Gopal, Ross Cutler, Eyal Krupka}}
\address{Microsoft Corporation \\
    \{ilyagu, idol, dharmendar.palle, yossiasher, alvinn, igorab, vishak.gopal, ross.cutler, eyalk\}@microsoft.com}

\begin{document}
\ninept %

\thispagestyle{firstpage}

\maketitle
\begin{abstract}
We introduce a distinctive real-time, causal, neural network-based active speaker detection system optimized for low-power edge computing. This system drives a virtual cinematography module and is deployed on a commercial device. The system uses data originating from a microphone array and a 360-degree camera. Our network requires only 127 MFLOPs per participant, for a meeting with 14 participants. Unlike previous work, we examine the error rate of our network when the computational budget is exhausted, and find that it exhibits graceful degradation, allowing the system to operate reasonably well even in this case. Departing from conventional DOA estimation approaches, our network learns to query the available acoustic data, considering the detected head locations. We train and evaluate our algorithm on a realistic meetings dataset featuring up to 14 participants in the same meeting, overlapped speech, and other challenging scenarios.
\end{abstract}


\section{Introduction}
\vspace{-5pt}
\label{sec:intro}
In the era of hybrid work environments, where teams combine in-person and remote collaboration, it is essential to ensure an equitable experience for all participants in meetings. This paper introduces and focuses on an active speaker detection (ASD) system running on a teleconferencing device placed on a table in a meeting room. The system's objective is to determine, in real-time, whether each participant in the meeting room is speaking or not. The ASD system then feeds into a virtual cinematographer that crops the speakers' faces, adjusts the virtual camera’s angles, and switches between different participants based on their speaking activity, creating a seamless and engaging visual experience for remote participants.

We present a novel real-time causal system that uses a horizontal circular microphone array in addition to a 360\degree camera to accurately determine who’s speaking in a meeting room. Our deep neural network runs on an edge device, powered by a lightweight Intel Movidius Myriad X vision processing unit (VPU) consuming only 2 Watts of power and that can concurrently handle up to 14 participants at a prediction rate of 7.5 predictions per second, requiring only $(123.5+43.5/K)$ MFLOPs to process a single participant’s head per frame, where K is the total number of participants in that frame. We also demonstrate that our algorithm exhibits graceful degradation when the number of participants exceeds the available computational budget.

While research that utilizes microphone array data is usually concerned with estimating the direction of arrival (DOA) of sounds, we present a distinctive approach for querying the available acoustic data given the location of the participant in question. Specifically, since we have video data available, we extract heads from it and then construct a representation that encodes the location of the participant we want to determine speech for. We also experiment with encoding the locations and sizes of the background participants as part of that query, to encourage the network to disregard possible interfering sources. This constitutes an end-to-end approach, rendering post-processing in the form of matching DOA estimates to participants’ locations unnecessary. Moreover, unlike previous audio-only methods which attempt to regress the azimuth direction only, our approach explicitly accounts for both azimuth and altitude information in the audio signal.
Another branch of our neural network models lip motion, which is correlated with the audio to accurately determine speech.

We train our system and evaluate it using an extensive and realistic dataset of multi-participant meetings collected specifically for this purpose, which features up to 14 participants in the same meeting, overlapped speech, and other challenging scenarios (see Sec. \ref{sec:dataset}). 

Our contributions include: (1) A real-time algorithm consisting of a head detector, a head tracker, an ASD deep neural network (DNN) model, and a virtual cinematography module running concurrently on a low-power edge device. (2) A low-latency neural network architecture that uses multi-channel audio and video feeds, in addition to spatial query data, to determine whether each participant is speaking. (3) A formulation of the ASD problem, which uses the participant's location as an input to the network to query the multi-channel audio data, and predict speech/silence class for it, in contrast to predicting a DOA. (4) Ablation studies to assess the contribution of input features and system components on the overall accuracy of the system. (5) A method to handle compute budget exhaustion and graceful degradation.

\mysection{Related work}
Over the past two decades, extensive research has been conducted on the ASD problem, starting with the pioneering work \cite{cutler_look_2000, zhang_boosting-based_2008}. This research can be characterized based on a whole range of factors, some of which we mention below.

\textit{Modalities:} Several modalities were employed for active speaker detection, each with its own advantages and limitations. Work utilizing visual modality (e.g.~\cite{guy2021learning}) relies on face detection and tracking techniques to feed classifiers that operate per face and detect facial cues, lip movements, or body language. However, relying solely on the video modality increases the likelihood of misinterpreting facial movements as certain actions are ambiguous. Furthermore, video is susceptible to limitations in cases such as unfavorable lighting conditions or when participants are occluded, face away from the camera, or located far from it.

When microphone array multi-channel audio is available, it becomes possible to utilize sound source localization (SSL) algorithms to determine the DOA of incoming sounds. Classical approaches to this problem include generalized cross-correlation-based methods \cite{knapp1976generalized, dibiase2000high}, and subspace-based methods. Lately, there has been a surge in the prevalence of DNN-based approaches to this problem \cite{grumiaux2022survey}. These include the direct usage of short-time Fourier transform (STFT)-based features, classical features, or their combinations. In spite of this considerable progress, SSL algorithms exhibit reduced robustness in situations where two or more speakers are situated in proximity, in cases of overlapped speech, or when background noise or reverberations are present.

A considerable amount of research has lately been focused on single-channel A/V fusion models. This challenging problem got much attention following \cite{chung2017out} and then \cite{roth_ava-activespeaker:_2020}, which introduced the AVA Active Speaker dataset. Most of that research builds on the premise that facial motion patterns can be correlated with the audio signal.

Few works were published lately that used multi-channel audio in combination with video. In \cite{jiang2022egocentric}, the authors use the audio signals to estimate a 2D heat map of acoustic activity in the scene and then concatenate it with a video frame while aligning the spatial coordinates. The result is then processed by another network to yield a final speech activity map. This work didn't use temporal modeling (which was shown to be beneficial for ASD \cite{tao2021someone}), and the audio pipeline made predictions without knowing the locations of the faces in the scene and without correlating lip movement with the audio. In \cite{qian2022audio}, audio-based features are combined with video-based features using a cross-modal attentive fusion mechanism proposed in the paper. However, video is used only to specify the locations of faces, without modeling information originating from the mouth region. The researchers of \cite{jiang2023incorporating}, whose work is most relevant to ours, proposed to incorporate lip features into a DOA estimation system. However, their system is not real-time, was trained on the MISP dataset which focuses on a scenario substantially different from ours (see discussion below), and requires postprocessing to associate DOA outputs to faces. Another difference from \cite{jiang2023incorporating} is that the latter detects the location of the lips, in addition to the face, thus spending compute time on this operation, as well as restricting the operational envelope of their system to near frontal poses.

Our model directly determines each participant's speech state, uses the knowledge of the locations of the heads to query the microphone array data, correlates lip movement with audio, and models long-term temporal relationships. It runs in real-time, doesn't require preprocessing in the form of detecting the lips, and is not limited to near frontal faces.

\textit{Inference run-time performance:} Despite significant progress made in the field of efficient neural networks (e.g., \cite{iandola2016squeezenet, howard2019searching, tan_efficientnetv2_2021}), the ASD problem received little attention from the community in this regard. Recently, \cite{liao2023light} proposed to split 3D convolutions into 2D and 1D convolutions to improve expressiveness and latency similar to the “(2+1)D” decomposition described in \cite{tran2018closer}. Their offline system, guided by this and other design choices, yielded near-SoTA results on AVA-ActiveSpeaker while requiring 200 MFLOPs per candidate per frame.

\textit{Datasets:} There are several datasets with some degree of relevance to our research. We list them below and outline the key factors that prompted us to eventually collect and conduct experiments with our own data. The \textit{AMI corpus} \cite{kraaij2005ami} includes multi-channel A/V recordings of meetings. However, there is a maximum of 4 participants per meeting. It also doesn't include audio-visual registration data, which prevents its use in algorithms in which it's needed. The \textit{MISP2021 challenge} \cite{chen2022first} and its accompanying dataset \cite{chen2022audio}, contains video and \textit{linear} microphone array recordings of up to 6 sitting participants (as described in \cite{jiang2023incorporating}), and focuses specifically on conversations in home TV rooms in Chinese, a scenario which is remarkedly different from ours. Unfortunately, we were also not able to experiment with it due to its restrictive license agreement. The popular single-channel A/V \textit{AVA Active Speaker} dataset \cite{roth_ava-activespeaker:_2020}, despite containing challenging scenarios such as dubbing and complex scenes, is highly skewed towards the film industry, and does not reflect the “true” distribution of human activity, as noted by its curators. The ASD task on this dataset is made simpler by taking advantage of priors relating to cinematographic effects (e.g., the camera tends to focus on the speaker \cite{zhang2021unicon}). Furthermore, in 90\% of the frames of this dataset, 3 or fewer participants are visible. This contrasts with our dataset, which contains between 4 to 7 people in 75\% of the frames, and 8 participants or more in 14\% of the frames.

\mysection{Method}

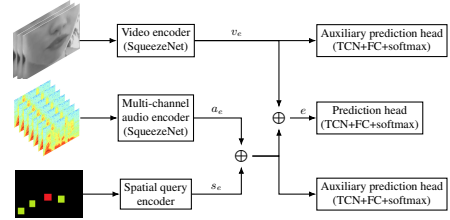
\begin{figure}
  \centering
  \input{architecture_horizontal}
  \caption{Proposed network architecture diagram}
  \label{arch}
\end{figure}

Our network architecture is illustrated in Fig.~\ref{arch}. Using three backbone networks, it first creates short-term representations of the A/V modalities and the spatial query information. We’ve chosen to use SqueezeNet \cite{iandola2016squeezenet} as a backbone network as it provides a good accuracy/performance tradeoff on the VPU. Then these embeddings are fed into a sequential model to be fused together, and to take long-term temporal context into account. For the sequential model, we’ve opted to select a TCN due to its low computational burden and superiority in a wide range of tasks \cite{bai2018empirical}.
At inference time, we take a sliding window approach: we maintain a first-in-first-out queue that contains the short-term embeddings. Whenever new embeddings become available, they're concatenated and put in a queue. If the length of the queue exceeds the receptive field size of the TCN the oldest item is removed. This simple approach allows the network to consider long-term information while calculating only the most recent embeddings at each timestep.

\mysubsection{Head detection and tracking}
The head detection and tracking module provides the locations of all persons in the room in each frame captured by the camera. The unconstrained meeting scenario involves many challenges, including occlusions, extreme head pose, varying lighting conditions, low resolution due to device-to-person distance, and motion blur. Therefore, any individual frame may not contain the necessary information for detecting all the people in the room. The head tracking uses head detection and low-level tracking to maintain a set of tracklets, where each tracklet is defined as a sequence of heads in time that belongs to the same person. We use a method similar to that in \cite{ren2008finding} with several adaptations to our specific setting, such as exploiting the stationarity of the camera for detecting motion, performing the low-level tracking by color-based mean-shift instead of gray-level based normalized correlation, tuning the algorithm to minimize the risk of tracklet mergers (which in our context are destructive), etc. Special attention was paid to meet the requirement for the real-time tracking of many people in the 360\degree panorama video. This was achieved by dividing the head detection and tracking task between two processes – one for searching for new heads to track and one for tracking the heads found by the former process. To avoid large latency, the detector of the first process is applied to each frame to part of the frame only, each time on a different part of it in a round-robin fashion. To increase the efficiency of the second process, motion detection is applied on all head regions being tracked. When there is no motion, the head location remains the same, so there is no need to apply tracking. Since motion detection is cheaper than tracking, and since most of the meeting heads are stationary, a lot of compute is saved.

\mysubsection{Visual encoder}
\label{sec:vis_enc}
We use the output of the tracker to crop the image patches of the participants, resize them to a fixed size $H \times W$, and convert them to grayscale. Prior to cropping, the bounding box undergoes a small adjustment in its location and size to make sure that the lip region is included in it when the face is visible (even when the participant is facing sideways). The parameters of this adjustment are found in an initial experimentation stage. These transformations of the bounding box allowed pixels that are more informative to be included in the network's receptive field. This simple approach makes the calculation of the landmarks (and specifically lips) unnecessary, thus saving computing time. To include movement information in the representation, we stack the last $l$ facial patches as channels to the backbone network $f_v$. Since $l$ is expected to be small, and since we smooth the output of the tracker before extracting the patches, the lack of registration between facial landmarks is minimal. Formally, we encode the short-term facial representation at each time-step $t$ for participant $i$, given its facial crop $v_{t,i} \in \mathbb{R}^{H \times W}$, using the backbone network $f_v$ as follows: $v_{t,i}^e = f_v(v_{t-l+1,i} \oplus v_{t-l+2,i} \oplus \ldots \oplus v_{t,i})$, where $\oplus$ denotes concatenation.

\mysubsection{Audio encoder}
Our system uses $T_a$ seconds, ending at frame $t$, of the $M$ microphones’ waveform data, sampled at frequency $\nu$ to encode the multi-channel audio signal $a_t\in\mathbb{R}^{T_a\nu\times M}$. STFT is then applied on each microphone’s $m\in\{1,\ldots,M\}$ signal $a_t[\cdot,m]$ separately. We then use the resulting spectrogram $S^m\in\mathbb{R}^{T_B\times F}$, where $T_B\times F$ are the dimensions of the time-frequency bin matrix, to extract simple log-magnitude and phase features, concatenate them along the channels dimension, and feed them into the audio backbone network $f_a$. Formally, the momentary audio representation is encoded as follows: $a_t^e=f_a\left(\bigoplus\limits_{m\in\{1,\textellipsis,M\}} log(\left|S^m\right|+\epsilon)\oplus arg(S^m)\right)$.

The result of the concatenation has the dimensions $2M\times T_B\times F$, where the first dimension indexes the norm and phase signals of the channels. This allows the 2D-CNN network to integrate data across all microphones for each time-frequency bin.

\mysubsection{Spatial query encoder}
We use the shared coordinate system of the 360\degree camera and the circular microphone array to encode a query containing spatial information of the reference and background participants. We first construct vectors $\mathbf{v}$ containing the sine and cosine of the azimuth $\lambda$ and the altitude $\phi$ of the participant’s head, the angular width of the head $\theta$, and the spherical distance (on a unit sphere) $\delta$ between the reference head and the current background head: $\mathbf{v} = \left(\sin{\lambda}, \cos{\lambda}, \sin{\phi}, \cos{\phi}, \theta, \delta\right)$.

We then sort background participants by their distance from the reference head and take $N$ background heads that are closest to the reference head. Then, we encode the reference head’s vector using a 2-layer fully connected (FC) network $f_{\text{ref}}$. A separate network with an identical architecture $f_{\text{bg}}$ is used to encode the background heads’ vectors. We then take the element-wise mean of these background vectors’ encodings, concatenate it with the encoding of the reference vector, and feed the result into another FC combiner network $f_{\text{comb}}$: $s_{t,i}^e = f_{\text{comb}}\left(f_{\text{ref}}\left(\mathbf{v}_{t,i}\right) \oplus \frac{1}{N}\sum_{j\in\{1,\ldots,N\}}{f_{\text{bg}}(\mathbf{v}_{t,j})}\right)$

This representation allows our network to reason about the spatial location of the reference participant, their distance from the device, and their interaction with possible distracting (background) participants.

\mysubsection{Fusion and temporal modeling}

We define a prediction head as $f_{\text{pred}}(v) = \text{softmax}(\text{FC}(TCN(v)_t))$. The fully connected network $\text{FC}$, which serves as a bottleneck, is 3 layers deep, and is applied on $TCN(v)_t$, the last timestep of the TCN’s output. It is followed by a 2-class softmax operation for speech/silence classification. The concatenation of all three encodings is defined as 
$e_{t,i} = v_{t,i}^e \oplus a_{t,i}^e \oplus s_{t,i}^e$. When training, a binary cross-entropy loss function $L_{\text{BCE}}$ is applied to the result, yielding the primary loss $L_{\text{vas}} = L_{\text{BCE}}\left(f_{\text{pred}}(e_{t-K+1,i}, \ldots, e_{t,i})\right)$, where $K$ is TCN’s receptive field. To encourage robustness to modality-specific corruptions and avoid cross-modality co-adaption, we apply auxiliary losses to separate groups of modalities $L_v = L_{\text{BCE}}\left(f_{\text{pred}}(v_{t-K+1,i}^e, \ldots, v_{t,i}^e)\right)$ and \\ $L_{\text{as}} = L_{\text{BCE}}\left(f_{\text{pred}}(a_{t-K+1,i}^e\oplus s_{t-K+1,i}^e,  \ldots, a_{t,i}^e\oplus s_{t,i}^e)\right).$
Note that the audio and the spatial query encodings must be concatenated together, as the latter contains the location of the reference participant. The final loss is $L = L_{\text{vas}} + \lambda_v L_v + \lambda_{\text{as}} L_{\text{as}}$.

\mysubsection{Data augmentation}
We augment the video data by randomly rotating and translating the facial clips to simulate head rotations and head detector's jitter, respectively. Audio is augmented by audio channel swapping, similarly to \cite{wang2023four}, which simulates the rotation of the microphone array by $\frac{360}{M} \cdot k$ degrees, where $k \in \{0, \ldots, M-1\}$, by taking advantage of its circular symmetry. The azimuth input to the spatial query encoder is rolled in tandem to maintain consistency.

\mysection{Experiments}

\mysubsection{Dataset}
\label{sec:dataset}
\begin{figure}  %
    \centering
    \includegraphics[width=0.5\columnwidth]{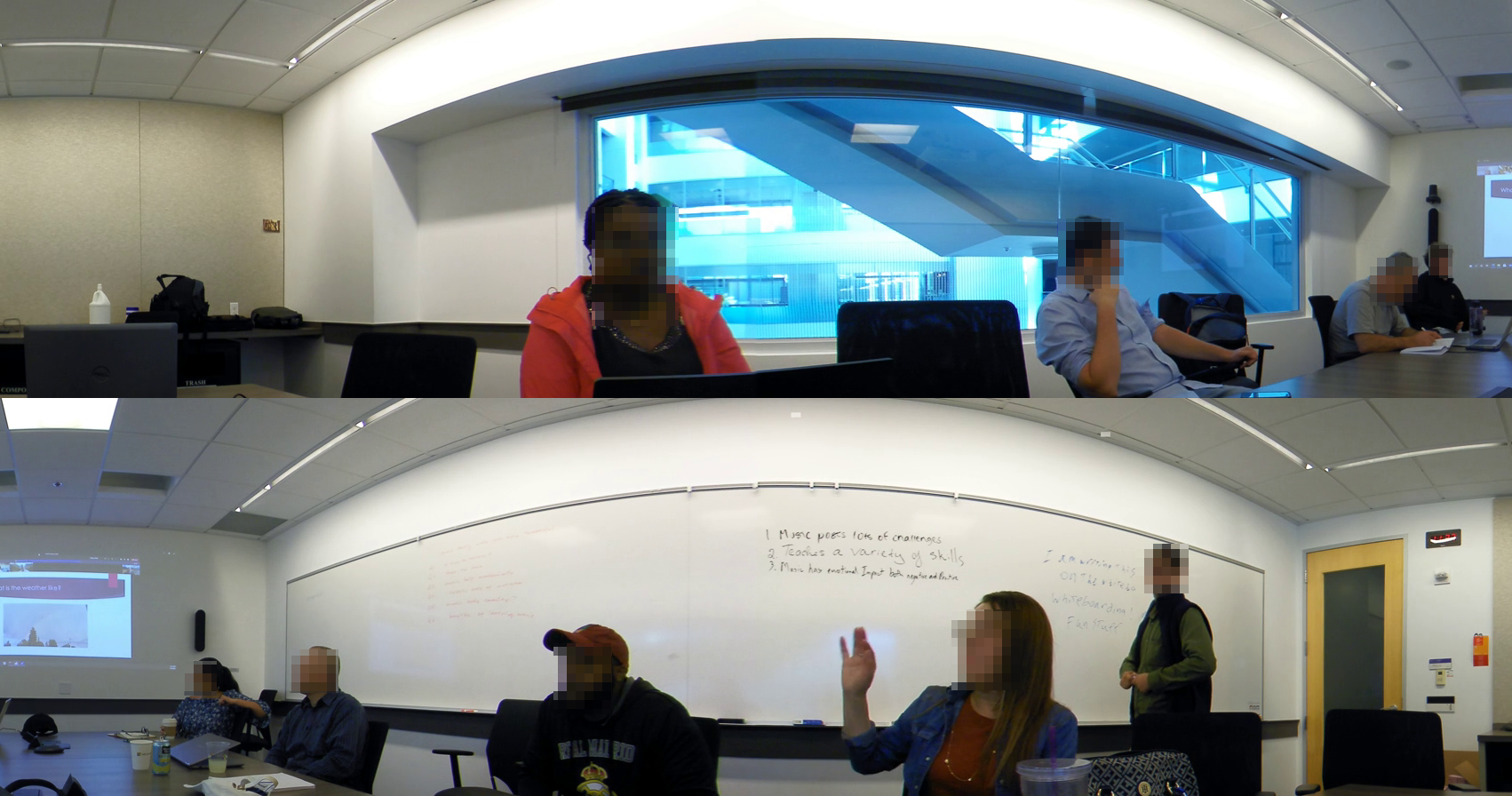}
    \caption{A frame from our dataset, arranged in two rows. The faces were pixelated to preserve participants' privacy.}
    \label{fig:enter-label}
\end{figure}

The commercial system we are building is intended to run on a device placed strategically in the center of a conference room table to enable capturing and transmitting the A/V signals of all meeting participants. The device’s camera is located approximately at the height of meeting participants’ faces similar to the RoundTable system \cite{zhang_boosting-based_2008}, has a full panoramic resolution of $1666 \times 10000$ pixels (height $\times$ width), and is accompanied by a circular microphone array with a radius of 4.25cm, which is situated slightly above the camera. The array comprises 6 MEMS microphones evenly spaced around its circumference, along with an additional microphone placed at its center.
We have collected 110 meetings, each 30 minutes long. We partitioned the dataset into 71, 17, and 22 meetings for the train, validation, and test sets, respectively. To avoid overfitting to specific participants, we made sure that every participant appeared in only one partition: 29, 17, and 14 participants in the train, validation, and test sets, respectively. The meetings were conducted in English; however, our participants came from diverse ethnic backgrounds and spoke with various accents. The dataset contains an equal proportion of males and females.  
The recorded meetings are not scripted but conducted in a natural manner so that each participant is free to speak and behave in a way that he feels appropriate. However, to kick-start each meeting we provide a topic for each discussion, e.g.: Pros/cons of working from home or office, etc. We also assign each participant a set of activities they need to perform such as whiteboarding, walking around, gesturing, entering/exiting the room, etc. Though these and other topics and activities can be raised and occur naturally, similarly to \cite{kraaij2005ami} we choose to explicitly elicit them to promote good coverage of real-world meetings.
The recordings are annotated at intervals of 200ms, specifying for each participant whether that participant is speaking or not. This rate is sufficient for our purpose of implementing a virtual cinematographer.

\mysubsection{Implementation details}
We reshape each patch to $120 \times 192$ pixels and set the number of frames $l$ (defined in Sec. \ref{sec:vis_enc}) for the short-term video representation to 3. Though our device captures video at a frame rate of 30 FPS, our ASD system operates at 7.5 FPS, as we found in initial experiments that this frame rate provides a good tradeoff between accuracy and compute time. The audio is sampled at a 16 kHz rate, and we use a window size of 512 samples, a hop length of 160 samples, and 512 FFT bins as STFT parameters. We set $T_a$, the duration of the waveform we process at each timestep, to 300 milliseconds. Following initial experimentation, we configured the TCNs to have three 64-channel layers. Our network architecture is implemented in PyTorch \cite{paszke2019pytorch}. For training, we use the SGD optimization algorithm with a Nesterov momentum of 0.9 and batch size of 64. Training is stopped if the validation error doesn't improve for 10 epochs.

\mysubsection{Evaluation metrics}
In SSL literature, the mean absolute DOA error is commonly used as an evaluation metric in contrast to binary classification metrics (e.g., F1, AUROC, etc.), which are used in the majority of ASD-related work. We find the latter to be more informative in scenarios such as ours, in which sparsely located candidates are first found in a preprocessing step. Object (and specifically: head) detectors don’t usually suffer from spatial error, reinforcing our point. 
In \cite{jiang2023incorporating} the authors calculated accuracy and F1 score by applying a threshold of 20\degree on the DOA error. In our scenario, however, we are interested in detecting the active speaker regardless of the azimuth difference from neighboring speakers. Thus, confusing who the active speaker is between two neighbors is considered an error even if they are at the same azimuth. We therefore calculate the equal error rate (EER) for all (frame, participant) pairs w.r.t.~the ground truth annotations.

\mysubsection{Ablation studies}

\begin{table}[]
\centering
\begin{tabular}{l c c c c c c l} \hline
    &\textbf{Vis}   & \textbf{Aud} & \textbf{Query} & \textbf{Bg} & \textbf{Aux} & \textbf{EER (\%)} \\ \hline
C1  & \checkmark    &              &                &                     &              &  9.31             \\
C2  & \checkmark    & \checkmark   &                &                     &  \checkmark  &  8.47             \\
C3  & \checkmark    & 1ch          &                &                     &              &  8.27             \\
C4  &               & \checkmark   &                &                     &              &  42.32            \\
C5  &               & \checkmark   & \checkmark     &                     &              &  7.36             \\
C6  &               & \checkmark   & \checkmark     & \checkmark          &              &  7.13             \\
C7  & \checkmark    & \checkmark   & \checkmark     &                     &  \checkmark  &  6.02             \\
C8  & \checkmark    & \checkmark   & \checkmark     & \checkmark          &  \checkmark  &  \textbf{5.99}    \\
C9 & \checkmark     & \checkmark   & \checkmark     & \checkmark          &              &  7.00             \\
\end{tabular}
\caption{Ablated configurations and results. Bg denotes background speakers' locations encoding. Aux denotes auxiliary loss.}
\label{ablation}
\end{table}

In Table \ref{ablation}, we show different system configurations, with certain features being either activated or deactivated for each one. Configuration C1, which uses only the visual signal, achieves an EER of 9.31\%, while C6, which uses the audio signal in combination with the full spatial query achieves an EER of 7.13\%. Their combination, C8 reduces the error rate to 5.99\%. However, when this A/V network is trained without auxiliary losses (C9) the error rate increases significantly to 7.00\%, emphasizing its importance.
Exploring the effects of ablating components of the query when only the audio signal is available, we find that excluding the background speakers’ representation increases the error slightly from 7.13\% (C6) to 7.36\% (C5). However, when both A/V signals are used, the network achieves similar error rates regardless of background speakers' encoding (C7, C8), probably by using information available in the visual modality to achieve similar results.
When the query is eliminated altogether (C2) the visual and the audio trunks work by correlating visemes with phonemes resulting in an EER of 8.47\%, which is better than the visual trunk alone (C1). The lack of any spatial information (query) obviates the need for multi-channel audio and indeed, a system that takes as inputs the visual signal with a single-channel audio signal (C3) achieves a similar accuracy. This result, together with previous ones, suggests that the audio trunk is needed not only to determine whether speech is coming from a certain direction but also to correlate lip motion with audio.
A system that uses the audio modality but is not provided with the query (C4) yields a very high error rate, as expected. 

\mysubsection{Inference time and graceful degradation}
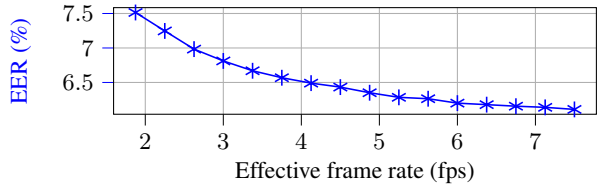
\begin{figure}
    \centering
    \input{graceful_degradation}
    \caption{EER versus frame rate}
    \label{fig:gradual_degradation}
\end{figure}

Our full network requires 167 MFLOPs (13ms on our target VPU device) to make a single inference for a single participant, out of which 43.5 MFLOPs (4.2ms) are spent on the audio encoder. As the audio encoder’s representation is participant independent, we reuse its result for all participants in a frame. Running at 7.5 predictions per second, our computational budget is 133.3ms, allowing us to support up to 14 participants in the room.
If this limit is exceeded, our system chooses the participants to predict in a round-robin fashion. That is, for every timestep the participants are sorted according to their last prediction times. Those that are the oldest get predicted till the computational budget is exhausted. This algorithm results in a stride that is different from the one that was used during training, constituting out-of-distribution data. Moreover, the stride between consecutive invocations may not be constant. We simulate round-robin in evaluation time and report the error rate as a function of the average prediction rate in Fig. \ref{fig:gradual_degradation}. We observe that the EER increases smoothly, up to 7.51\% at an average prediction rate of 1.875 predictions/sec, which is still a reasonable figure for our purposes, especially given the fact that the model was not trained to handle this scenario.

\mysection{Conclusions}
We have described a real-time efficient ASD system whose performance gracefully degrades under heavy computational constraints. We have departed from traditional DOA estimation methods by querying available acoustic data using detected head locations, eliminating the need for post-processing stages, thus taking another step towards end-to-end architectures for multi-channel ASD.

\vfill\pagebreak

\bibliographystyle{IEEEbib}
\bibliography{IC3-AI,refs}

\end{document}

%% file: architecture_horizontal.tex
\tikzset{
  myarrow/.style={
    -{Latex[length=1mm]}
  }
}

\begin{tikzpicture}[scale=0.5, transform shape,
  STY/.style={rectangle, align=center, draw=black}
]

  \node[STY] (vis) {Video encoder\\(SqueezeNet)};
  \node[STY, below=of vis] (mca) {Multi-channel\\audio encoder\\(SqueezeNet)};
  \node[STY, below=of mca] (ctx) {Spatial query\\encoder};
  
  \node[right=of $(mca.east)!0.5!(ctx.east)$] (cc_part) {$\bigoplus$};
  \node[right=of cc_part.west|-mca.east] (cc_all) {$\bigoplus$};

  \node[inner sep=0pt, left=of vis] (vis_im) {\includegraphics[width=0.10\textwidth]{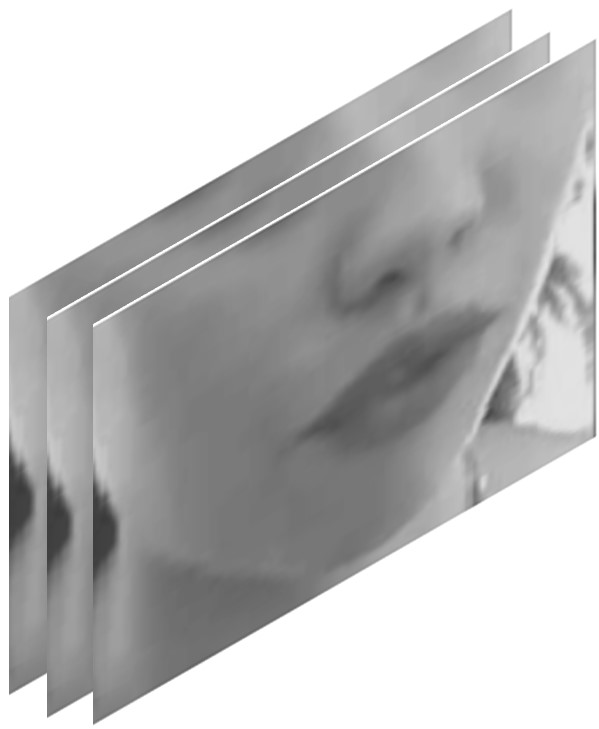}};
  \node[inner sep=0pt, left=of mca] (mca_im) {\includegraphics[width=0.10\textwidth]{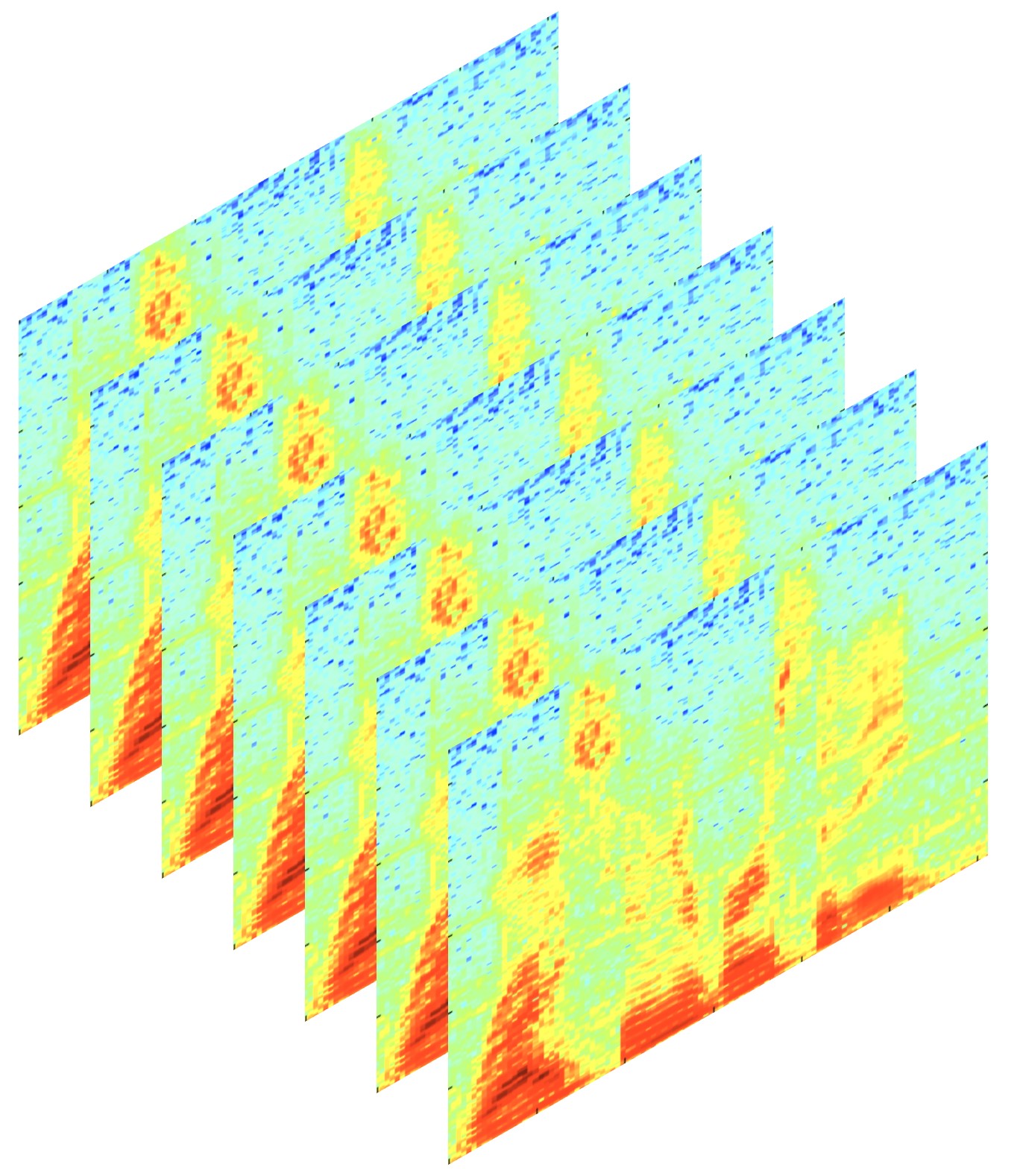}};
  \node[inner sep=0pt, left=of ctx] (ctx_im) {\includegraphics[width=0.10\textwidth]{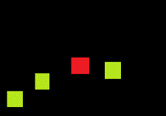}};

  \node[STY, right=of cc_all|-vis] (vis_aux) {Auxiliary prediction head\\(TCN+FC+softmax)};
  \node[STY, right=of cc_all|-mca] (temp) {Prediction head\\(TCN+FC+softmax)};
  \node[STY, right=of cc_all|-ctx] (part_aux) {Auxiliary prediction head\\(TCN+FC+softmax)};

  \draw[myarrow] (vis_im.east) -- (vis.west);
  \draw[myarrow] (mca_im.east) -- (mca.west);
  \draw[myarrow] (ctx_im.east) -- (ctx.west);
  
  \draw[myarrow] (vis.east) to node[pos=0.5, above] {$v_e$} (cc_all.north|-vis.east) -- (cc_all.north);
  
  \draw[myarrow] (vis.east) -- (vis_aux);
  \draw[myarrow] (cc_part.east) to (cc_all.south|-cc_part.west) -- (cc_all.south|-ctx.west) --(part_aux.west);

  \draw[myarrow] (mca.east) to node[pos=0.5, above] {$a_e$} (cc_part.north|-mca.east) -- (cc_part.north);

  \draw[myarrow] (ctx.east) to node [pos=0.5, above] {$s_e$} (cc_part.south|-ctx.east) -- (cc_part.south);

  \draw[myarrow] (cc_part.east) to (cc_all.south|-cc_part.west) -- (cc_all.south);

  \draw[myarrow] (cc_all) -- node[pos=0.5, above] {$e$} (temp);

\end{tikzpicture}

%% file: graceful_degradation.tex
\begin{tikzpicture}

\definecolor{darkgray176}{RGB}{176,176,176}

\begin{axis}[
tick align=outside,
tick pos=left,
x grid style={darkgray176},
xlabel={Effective frame rate (fps)},
xmajorgrids,
xmin=1.59374999999999, xmax=7.78125,
xminorgrids,
xtick style={color=black},
y grid style={darkgray176},
ylabel=\textcolor{blue}{EER (\%)},
ymajorgrids,
ymin=6.03901138907187, ymax=7.58480881392437,
yminorgrids,
ytick style={color=blue},
height=3cm,
width=8cm
]
\addplot [semithick, blue, mark=asterisk, mark size=3, mark options={solid}]
table {%
7.5 6.10927490838334
7.125 6.13915845243686
6.75 6.15578538672227
6.375 6.17825421683769
6 6.2001281745184
5.625 6.26421543371996
5.25 6.28432124052828
4.875 6.35092172558087
4.5 6.4326015657397
4.125 6.49012157883975
3.75 6.56853779594258
3.375 6.66964753146198
3 6.81209866924691
2.625 6.98353708817443
2.25 7.24709646542833
1.875 7.5145452946129
};
\end{axis}

\end{tikzpicture}